\def   \ni {\noindent}

\def   \ssk {\vskip  5truept}

\def   \bsk {\vskip 15truept}
 
\def   \newpage {\vfill\eject}
\def   \newline {\hfil\break}

\def\and{{\&}}

\def\dex1{\mbox{dex}}
\def\dex{\hbox{\rm dex}}

\def\eg1{{e.g.\/}}

\def\gp{\hbox{\rlap{\hbox{.}}\raise 5.truept \hbox{{\small$\circ$}}}}
\def\gradip{\hbox{\rlap{\hbox{.}}\raise 5.truept \hbox{{\small $\circ$}}}}
\def\orep{\hbox{\rlap{\hbox{.}}\raise 5.truept \hbox{{\small $h$}}}}

\def\magcir{\ \raise-2.truept\hbox{\rlap{\hbox{$\sim$}}\raise5.truept
\hbox{$>$}\ }}

\def\mincir{\ \raise-2.truept\hbox{\rlap{\hbox{$\sim$}}\raise5.truept
\hbox{$<$}\ }}

\def\underline{}

\def\lsim{\,\lower2truept\hbox{${< \atop\hbox{\raise4truept\hbox{$\sim$}}}$}\,}
\def\gsim{\,\lower2truept\hbox{${> \atop\hbox{\raise4truept\hbox{$\sim$}}}$}\,}

\def\s{\,{\rm s}}

\def\s_fit{\mbox{$\sigma_{\rm bf}$}}

\def\dex{\mbox{dex}}

\def\eg{{\it e.g}}

\def\$\sigma$c{\mbox{$\$\sigma$_c$}}

\def\gp{\hbox{\rlap{\hbox{.}}\raise 5.truept \hbox{{\small$\circ$}}}}

\def\ni{\noindent}                          

\def\lsim{\, \lower2truept\hbox{${< \atop\hbox{\raise4truept\hbox{$\sim$}}}$}\,}
\def\gsim{\, \lower2truept\hbox{${> \atop\hbox{\raise4truept\hbox{$\sim$}}}$}\,}
\documentstyle[epsfig]{article}
\begin{document}

%
\def\la{\mathrel{\mathchoice {\vcenter{\offinterlineskip\halign{\hfil
$\displaystyle##$\hfil\cr<\cr\sim\cr}}}
{\vcenter{\offinterlineskip\halign{\hfil$\textstyle##$\hfil\cr
<\cr\sim\cr}}}
{\vcenter{\offinterlineskip\halign{\hfil$\scriptstyle##$\hfil\cr
<\cr\sim\cr}}}
{\vcenter{\offinterlineskip\halign{\hfil$\scriptscriptstyle##$\hfil\cr
<\cr\sim\cr}}}}}
\def\ga{\mathrel{\mathchoice {\vcenter{\offinterlineskip\halign{\hfil
$\displaystyle##$\hfil\cr>\cr\sim\cr}}}
{\vcenter{\offinterlineskip\halign{\hfil$\textstyle##$\hfil\cr
>\cr\sim\cr}}}
{\vcenter{\offinterlineskip\halign{\hfil$\scriptstyle##$\hfil\cr
>\cr\sim\cr}}}
{\vcenter{\offinterlineskip\halign{\hfil$\scriptscriptstyle##$\hfil\cr
>\cr\sim\cr}}}}}
\def\degr{\hbox{$^\circ$}}
\def\arcmin{\hbox{$^\prime$}}
\def\arcsec{\hbox{$^{\prime\prime}$}}

\hsize 5truein
\vsize 8truein
\font\abstract=cmr8
\font\keywords=cmr8
\font\caption=cmr8
\font\references=cmr8
\font\text=cmr10
\font\affiliation=cmssi10
\font\author=cmss10
\font\mc=cmss8
\font\title=cmssbx10 scaled\magstep2
\font\alcit=cmti7 scaled\magstephalf
\font\alcin=cmr6 
\font\ita=cmti8
\font\mma=cmr8
\def\ref{\par\noindent\hangindent 15pt}
\null


\title{\ni 
The Planck Low Frequency Instrument}

\bsk 
\author{\ni N.~Mandolesi$^1$, M.~Bersanelli$^2$, C.~Burigana$^1$ \& F.~Villa$^1$
\\
On behalf of LFI Consortium
} 
\bsk
\affiliation{$^1$Istituto TeSRE, CNR, Bologna, Italy; 
$^2$IFC, CNR, Milano, Italy.}
\bsk
\baselineskip = 9pt

\abstract{ABSTRACT -- \ni
The Low Frequency Instrument (LFI)  of the ``Planck Surveyor"
ESA mission will perform high-resolution imaging of the Cosmic Microwave
Background anisotropies at four frequencies in the 30--100 GHz range. We
review the LFI main scientific objectives, the current status of the
instrument design and the on-going effort to develop software simulations
of the LFI observations. In particular we discuss the design status of the
PLANCK telescope, which is critical for reaching adequate effective
angular resolution. 
}                                                    
\bsk
\baselineskip = 9pt

\bsk
\baselineskip = 12pt


\text{\ni 1. INTRODUCTION
\ssk
\ni     
The Planck LFI represents the third generation of mm-wave instruments
designed for space observations of CMB anisotropies, following the COBE
Differential Microwave Radiometer (DMR) and the Microwave Anisotropy Probe
(MAP). The DMR, launched in 1989, detected structure in the CMB angular
distribution at angular scales $>7\deg$.  The LFI will produce images of
the sky at four frequencies between 30 and $100\,$GHz, with an
unprecedented combination of sky coverage, calibration accuracy, freedom
from systematic errors, stability and sensitivity (including polarized
components). The LFI will produce full
sky maps at 30, 44, 70 and 100 GHz, with angular resolution of $33'$,
$23'$, $14'$ and $10'$, respectively, and with an average sensitivity per
resolution element $\Delta T/T \simeq$ a $few \times 10^{-6}$.
These unprecedented angular resolution and sensitivity will
uncover the wealth of cosmological information encoded in the anisotropy
pattern at degree and sub-degree angular scales.

In the LFI frequency range the contaminating
effect of the galactic emission, dominating the astrophysical foreground
noise on scale $\geq 30'$, is minimum at around $60\,$GHz; while the
confusion noise due to extragalactic sources, dominating on smaller
angular scales, is minimum in the range 100--$200\,$GHz, where it is
primarily due to radio sources
(De Zotti \& Toffolatti 1998). The 70 and $100\,$GHz channels are
therefore optimal to get the cleanest possible view of primordial CMB
fluctuations, over the full range of angular scales. In both channels the
astrophysical foreground noise is expected to be well below the
cosmological signal for all observed angular scales.  At $100\,$GHz then
the LFI will accurately measure the power spectrum of CMB anisotropies up
to multipoles $\ell \sim 1300$ with an accuracy of the order of, or
better, than 1\%. Little cosmological information is left at angular
scales smaller than 10 arcminutes, if standard inflationary models hold;
in fact anisotropies at such scales are quasi-exponentially erased by
photon diffusion.

\newpage

The LFI measurements will determine of the primary
cosmological parameters (Hubble constant, deceleration parameter, 
curvature of space, baryon density, dark matter densities including
neutrinos, amplitude and spectral index of the primordial scalar density 
perturbations, and the gravity wave content of the Universe)
to an accuracy of a few percent (see, e.g., Bond et~al.~1997).
The LFI data can test models for the origin of primordial perturbations, i.e. 
whether they are due to topological defects or to quantum fluctuations, 
and constrain the global properties of the Universe 
(topology, rotation, shear, etc.)
and the theories of particle physics at energies $\simeq 10^{16}$~GeV.
Polarization measurements, which will be possible with an accuracy of 
a few $\mu$K towards the ecliptic caps, will independently confirm
these findings and help 
in breaking degeneracy in the determination of cosmological 
parameters (Zaldarriaga et~al.~1997).
Also, contraints on (or possible detections of) deviations of the 
CMB spectrum from a planckian shape can be accurately studied by Planck,
by analyzing the dipole signature, so providing interesting information
on cosmological and astrophysical processes at high redshifts 
(Danese \& De~Zotti 1981, Burigana et~al.~1998b). 

The LFI will also detect the Sunyaev-Zeldovich effect 
(Sunyaev \& Zeldovich 1970) towards a few 
hundred of clusters of galaxies, allowing an independent determination 
of the Hubble constant (Cavaliere et~al.~1979; Myers et~al.~1997) 
and providing information on the intercluster medium 
complementary to those from X-rays (Rephaeli 1995).
The LFI four-frequency all-sky surveys will also 
be unique in providing complete samples 
comprising from several hundred to a few thousands extragalactic sources, 
selected in an essentially unexplored frequency range,
like familiar ``flat-spectrum" radiosources, sources 
with strongly inverted spectra and possible new classes of sources
(Toffolatti et~al.~1998, De~Zotti \& Toffolatti 1998). 

Moreover, the LFI maps will provide a rich database for studies of
Galactic evolution, the interstellar medium, and discrete Galactic
sources, including supernova remnants, Sagittarius A, and sources
self-absorbed up to high frequencies such as some symbiotic stars and
planetary nebulae.

The combination of data from its two instruments, LFI and HFI (High
Frequency Instrument, see Puget et~al.~1998), 
give to the Planck Surveyor the imaging power, the redundancy and the control 
of systematic effects and foreground emissions needed to achieve the 
extraordinarily exciting scientific goals of this mission in a broad
spectral range, from 30 to 857$\,$GHz. 
This, in turn, is crucial for improving the accuracy in the determination 
of the cosmological parameters. 
LFI and HFI will deal primarily with different astrophysical 
processes, radio and dust emission, respectively, 
both coexisting in real astrophysical sources. 
The LFI data are crucial to separate the cosmic signal from the 
contaminating effect of extragalactic radio sources which dominate 
the foreground fluctuations on small angular scales in the 
cosmologically cleanest frequency range, at least up to 200 GHz. 
On the other hand, the HFI maps will be useful to subtract the 
Galactic dust emission which is important on intermediate to large angular 
scales down to $100\,$GHz. 
Also, the full set of Planck  data will be essential 
to address a number of important astrophysical problems such as to elucidate 
physical and evolutionary connections 
between nuclear activity (responsible for the radio emission) 
and processes governing the abundance and the properties 
of the interstellar material (responsible for the sub-mm dust emission). 

\bsk \ni 2. PROGRAMMATICS \ssk \ni Planck was formerly called COBRAS/SAMBA
(Bersanelli et~al.~1996), a combination of the two CMB proposals ``COBRAS"
and ``SAMBA"  submitted to ESA in 1993 in response to the call for mission
ideas for the Medium-Size M3 mission, expected to be launched in 2003. 
After the Assessment Study and Phase A Study, COBRAS/SAMBA was selected
and approved in late 1996, it was renamed in honor of Max Planck and the
launch was then planned for 2004.  The ESA Announcement Opportunity (AO)
for the instruments for the FIRST/Planck Programme was issued in 1997
announcing a launch in 2006.  Budgetary pressures within ESA's scientific
programme have forced a reconsideration of the original implementation
plan for Planck. Between 1997 and 1998 several studies were carried out to
determine how Planck will be implemented.  After we replied to the AO the
launch date was shifted to 2007.  The preferred option at the present time
is to launch Planck together with the FIRST mission; in this solution
(known as the "Carrier" configuration because of the launch arrangement)
both Planck
and FIRST will be placed in separate orbits around the second 
Lagrangian point of the Earth-Sun System.
The LFI data will be transmitted to the LFI Data Processing Center
(see Pasian \& Gispert 1998 for a detailed description).

\bsk
\ni 3. INSTRUMENT DESCRIPTION
\ssk
\ni 
A schematic overview of the LFI front-end unit is shown in Figure 1.  The
front-end unit is located at the focus of the Planck off-axis telescope in
a circular configuration around the High Frequency Instrument.  The
front-end unit is the heart of the instrument, and includes 28 modules,
each containing one feed horn, one orthomode transducer (OMT), two hybrid
couplers, two phase switches, and four cryogenic amplifiers.

\begin{figure}[h]
\centerline{\psfig{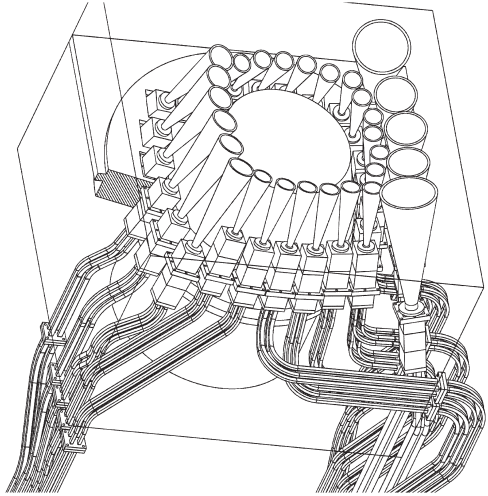}}
\caption{FIGURE 1. 
Schematic of LFI front-end unit in the asymmetric configuration.
}
\end{figure}

The radiation focused by the telescope is coupled to the radiometers by
conical, corrugated feedhorns (Bersanelli et~al.~1998, Villa et~al.~1997,1998a).
The radiation patterns
of the horns must be highly symmetric, with low sidelobes and a beam width
that matches the telescope edge taper requirement ($<-30$ dB at 22$\deg$). 
In addition, the electromagnetic field inside the horn must propagate with
low attenuation and low return loss. The OMTs separate the orthogonal
polarizations with minimal losses and cross-talk.  The return loss ($<-25$
dB) and the insertion loss ($<0.1$ to 0.3 dB depending on frequency) must
be met over the whole 20\% bandwidth. 

Each OMT is coupled to an integrated front-end module containing the
hybrid couplers (two for each module) and the amplifier chains including
phase switches and output hybrids. The front-end modules are mounted on
the 20\,K plate. Each hybrid has two inputs, one of which sees the sky
through one of the OMT arms and the feed horn, the other of which looks at
the 4\,K reference load through a small horn fabricated into the hybrid
block.  Four amplifiers are contained in each amplifier block, with
multiple inputs and outputs on a single flange to minimize size.  The
hybrid coupler combines the signals from the sky and cold load with a
fixed phase offset of either $90^{\circ}$ or $180^{\circ}$ between them. 
More details on the LFI receiver concept and potential systematics are
given by Bersanelli \& Mandolesi (1998).

The LFI Front End Unit is cooled to 20\,K by a hydrogen sorption cooler
(Wade \& Levy 1997). In addition to meeting temperature and capacity
requirements, the sorption cooler has major advantages for Planck:  it is
highly efficient; there are no cold moving parts, no vibration, and very
low EMI; and integration with the instruments and spacecraft is simple,
since the only part of the cooler in the focal assembly is a J-T valve and
associated tubing while the compressor and control electronics are located
remotely on the spacecraft.  The sorption cooler also provides 18\,K
precooling to the HFI helium J-T cooler.  The Planck thermal design allows
efficient radiative cooling of the payload and an optimized design of the
combined LFI and HFI front-end assemblies in the focal plane. 

Following amplification, the signals are passed through a cryogenic,
low-power phase switch, which adds $90^{\circ}$ or $180^{\circ}$ of phase
lag to the signals thus selecting the input source as either the sky or
the reference load at the radiometer output at a rate of about 1\,kHz. The
phase lagged pair of signals is then passed into a second hybrid coupler,
separating the signals.  The inclusion of the phase switches and second
hybrids in the front-end blocks eliminates the need for phase matching of
the long transmission paths to the back-end, greatly simplifying
integration and testing. 

Each cryogenic front-end module is connected with the room temperature
section with four waveguides or cables, grouped together. This results in
a total of 20 coax cables and 92 waveguide sections running between the
front-end unit and the back-end unit.  The transmission lines are further
grouped into four bundles and configured to allow the required flexibility
and clearance for integration of HFI in the central portion of the
assembly. 

Each back-end module comprises two parallel chains of amplification,
filtering, detection, and integration. The detected signals are amplified
and a low-pass filter or integrator reduces the variance of the random
signal, providing in each channel a DC output voltage related to the
average value.  Post-detection amplifiers are integrated into the modules
to avoid data transmission problems between the radiometer and the
electronics box.  The sky and reference signals are at different levels,
which are equalized after detection and integration by modulating the gain
synchronously with the phase switch. 

The thermal design is based on three principles:  a) minimize the power
dissipated in the focal assembly; b) maximize the effectiveness of
radiative cooling by providing good views of cold space, and by
intercepting conductive heat loads and radiating them away; c)  segregate
the warm and cold components, and keep them as far from each other as
possible.

\bsk
\ni 4. MISSION PERFORMANCE AND SIMULATIONS 
\ssk
\ni 

A large set of detailed simulation codes is being developed by both LFI
and HFI Consortia, in close contact with the theoretical and hardware
progress, with the aim of testing and contributing to improve the 
mission design.  We briefly summarize here the basic concepts on several
issues investigated through simulations.
 
A crucial effect is introduced by straylight, i.e., contamination from
off-axis sources through the sidelobes of the instrument beam.  
Several sources of contamination, both instrumental and astrophysical,
may introduce spurious signals.  The primary environmental sources of
error for the LFI are those due to imperfect off-axis rejection by the
optical system of radiation from the Sun, Earth, Moon planets, and Galaxy.
Sidelobe structure sweeping across the Galaxy can
produce artifacts in any direction. Sidelobe contributions are dominated
by the features relatively near the optical axis and typical maximum
ratios between the sidelobe and the central beam signals reach levels of
$\simeq 10\%$ (Mandolesi et~al.~1998, section 2.1.2);  we require the
level of galactic contamination be below the noise level with a factor of
two margin to allow for uncertainties in the level of galactic emission.
The exact contamination levels depend on the details of the sidelobe
pattern, typically in the range $-60$ to $-70$~dB (for a more detailed
study see Polegre et~al.~1998).  In addition, emission
from near field objects may affect the anisotropy
measurements, such as emission from the warm parts of the
spacecraft or fluctuations
of mirror and shield temperature.

Transistor gain fluctuations in the receivers introduce amplifier noise
temperature fluctuations that dominate in generating instrumental drifts
with a $1/f$ noise spectrum (Seiffert et~al.~1997). The LFI radiometer
design minimizes this effect, but residual stripes may be present.
Destriping algorithms (Delabrouille 1998, Burigana et~al.~1997), based on
the idea that the same position in the sky must give the same observed
temperature for different satellite spin axis positions, are quite
efficient in further reducing these stripes, provided that the crossings
between pointing positions in the sky obtained by different satellite spin
axis positions are spreaded enough in the sky. In terms of increased noise
with respect to the case of receiver pure white noise, these destriping
techniques allow to reduce the noise added by stripes from some tens
percent to few percent.  Even in the case in which the angle, $\alpha$,
between the telescope optical axis and the satellite spin axis is
constantly keeped at $90^{\circ}$, the worst from this point of view, for
the major part of LFI beams located at $2^{\circ} \div 4^{\circ}$ from the
telescope optical axis -- but not close to the sky scanning direction --
we find a good destriping efficiency. For on-axis beams (or, equivalently,
for beams located close to the sky scanning direction), reducing the angle
$\alpha$ significantly improves the efficiency of these destriping
methods.  By working with maps at resolutions close to the Planck
FWHM's, we find that an angle of $85^{\circ}$ is a good compromise for
having efficient destriping and, considering the spread of the
projected beam positions on the sky, practically full sky coverage.

Stripes in the observed maps can be also introduced by thermal
instabilities.  The closer the spin axis remains to the Sun-Planck
direction, the smaller are any temperature variation induced by departure
from perfect cylindrical symmetry.  Thermal design aims at producing both
low temperatures for sensitivity and thermal stability for reducing
drifts. Destriping algorithms can be applied also for reducing residual
stripes due to thermal instabilities, which typically show a noise
spectrum close to $1/f^2$ (Delabrouille 1998). 

The amplitude and the reduction in the data analysis of stripes introduced
by thermal and gain fluctuations is related to the Planck observational
strategy.  Sinusoidal oscillations of the spin axis may produce
significant
variations of the illumination by the Sun, so introducing unwanted thermal
instabilities.  On the other hand, different kinds of spin axis
``oscillations" in which the angle between the spin axis and the
Sun-Planck direction is keeped constant (like for example a precession
motion of the spin axis around an axis constantly keeped on the ecliptic
plane) minimise this effect and can be useful in reducing the
stripes even for on-axis beams in the case $\alpha = 90^{\circ}$. 

On the other hand, the scanning strategy controls another important issue:
only when $\alpha$ is constant the distribution in the sky of the
sensitivity per pixel is smooth, the global integration time per pixel
increasing continuously from the ecliptic equators to the ecliptic poles.
In the other cases, we can have large areas in sky where the sensitivity
significantly varies from a pixel to another even for small changes of the
position in the sky. This is exactly what we want to avoid to not
complicate the data analysis, particularly in presence of foreground
contamination. 

As discussed in the introduction, the most important LFI channels from the
cosmological point of view are at 70 and 100~GHz; in particular the
100~GHz channel presents the best LFI resolution. In the new symmetric
configuration of the Focal Plane Unit (FPU) the LFI feeds are located in a
ring around the telescope optical axis at about $2^{\circ} \div 4^{\circ}$
from it.  Therefore, the issue of the main beam optical distortions is
crucial, particularly at 100~GHz, the LFI channel where we want to reach a
FWHM angular resolution of $\simeq 10'$, necessary for the cosmological
goal, and where the optical aberrations are maximum, because of their
increasing with the frequency. 

\bsk
\ni 5. TELESCOPE DESIGN 
\ssk
\ni 
The optimization of the Planck telescope is one of the goals of the 
Planck Teams. 
For the present optical study 
we have considered the 100~GHz channel.

The baseline design (report on Phase-A, TICRA yellow report, 
etc..) is a 1292.4 mm 
projected aperture gregorian off-axis telescope satisfying
the Dragone-Mizuguchi condition (Dragone 1978; Mizuguchi et~al.~1978; 
Rush et~al.~1990). 
This condition set the tilting of the 
subreflector axis with respect the main reflector axis in order to cancel 
the cross-polarization. Unfortunately, only
in the center of the focal surface (null scan angle)
this kind of design shows symmetrical beams. 
Beam aberrations (expecially coma aberration) 
rise when the feedhorn is located outside the center of the 
focal surface, and increase with the frequency and the distance from the 
optical axis. Since one of the most crucial effect of beam 
distortions is an angular resolution degradation with respect to the central beam,
we have studied two configurations with increased main reflector apertures
in order to improve the resolution of the beams.
The first one has a projected aperture
of 1492.4 mm, while the projected aperture of the second configuration is 
1750.0 mm. All these configurations have the same subreflector of the 
1292.4 mm baseline design, as well as the same overall focal ratio. This means that
the angular geometry is preserved for all the designs 
and no relevant changes of the
FPU arrangement are needed.

Among other possible 
design configurations, an alternative to the Dragone-Mizuguchi 
Gregorian off-axis solution (in short ``Standard'') is represented by the 
Aplanatic Gregorian design (in short ``Aplanatic''), 
firstly proposed by Mark Dragovan and the LFI Consortium
in order to reduce the coma and 
the spherical aberration on
a large portion of the focal surface (Villa et~al.~1998b).
This new solution is obtained by changing
the conic constants of both mirrors (both ellipsoids of
rotation) in order to satisfy the Aplanatic Condition. 
Two configurations
have been studied, with 1292.4 mm and 1492.4 mm projected aperture 
respectively. 
Details of all the considered configurations, sketched in Figure 2, 
are reported in Table 1. 

\begin{figure}[h]
\centerline{\psfig{file=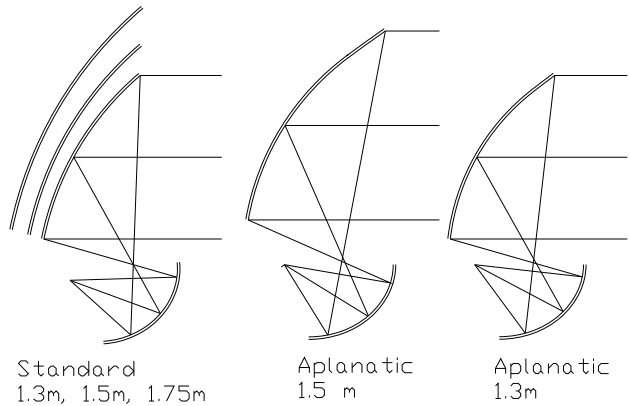, width=9cm}}
\caption{FIGURE 2. Schematic of the considered optical designs.}
\end{figure}

To analyze a general dual reflector system a dedicated software has been 
implemented at Istituto TeSRE/CNR (Villa et~al.~1998b). 
The code calculates the amplitude and phase distribution on a regular 
grid of points on the tilted aperture plane (normal to azimuth and
elevation directions on the sky).

The amplitude is calculated starting from the feed pattern and takes into
account the space attenuation.

\begin{table}
\begin{center}
\caption{TABLE 1. Characteristics of the Telescopes: The first three
configurations are Standard design. The last two are referred to the
Aplanatic design.}
\begin{tabular}{l | lll | ll}
\hline\hline
Main Refl. Diameter (mm)& $1292.4$ & $1492.4$ & $1750.0$ & $1292.4$ &
$1492.4$ \\ 
Main Refl. Shape & Par.&Par.&Par.&Ell. &
Ell.\\
Sub Refl. Tilting & $14^\circ$ & $14^\circ$  & $14^\circ$ & $0^\circ$&
$0^\circ$\\
Sub Refl. Focal Length (mm)& $514.29$ & $514.29$ & $514.29$ & $680.912$ &
$688.858$ \\
Sub Refl. Shape & Ell. & Ell. & Ell. & Ell. &Ell. \\
Overall Focal Ratio              & $1.39$ &$1.39$ &$1.39$ &$1.39$ &$1.39$
\\ 
\hline
\end{tabular}
\end{center}
\label{tb_ch}
\end{table}

The phase is derived by calculating the path length of each ray, from
the corresponding point on the aperture plane grid up to the focal point
previously calculated (minimizing the wave front error of the spot
diagram). Performing the Fourier Transform of the spatial phase-amplitude
distribution on the aperture plane, the far field radiation pattern is 
readily obtained. 
For each configurations the contour plot of the normalized patterns as function
of the sky-pointing scan angles (elevation, azimuth) have been calculated. 
Figure 3 shows our results for a typical beam position. 
Diffraction effects on the reflectors rim are not considered, but for the main 
beam response they are expected to be quite small.

\begin{figure}[h]
\centerline{\psfig{file=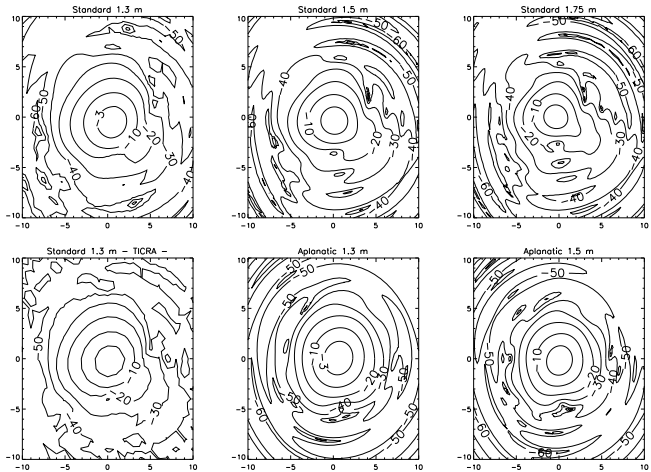, width=9cm}}
\caption{FIGURE 3. Contour plot of the normalized patterns of a beam 
located at El=$+2^{\circ}$, Az=$-2^{\circ}$ from the optical axis
for the considered optical designs; also the pattern computed by TICRA 
for the 1.3~m standard telescope is shown for comparison.
In each panel, the $x$ and $y$ axes are the standard U and V 
coordinates in radians (multiplied by
1000) in a reference roughly translated to the beam center.}
\end{figure}

All simulations have been done by considering the $\cos^N(\theta)$ primary
pattern with $N = 91$ (for the Standard 1.3 m configuration this gives an 
edge taper of $-30$ dB at $22^\circ$ of angle). 

In order to quantify the impact on the effective angular resolution,
FWHM$_{\rm eff}$, of the beams in CMB anisotropy measurements we have
compared 
convolutions of a CDM anisotropy sky with the simulated beams
and with a suitable grid of gaussian symmetric beams
(see Burigana et al. 1998a and Mandolesi et al. 1997, section 3.2, for
further details on the method). 
In Figure 4 we summarize our results for the five considered configurations.
Note that the August 1998 ESA Baseline Planck telescope is a 1492.4 m aperture 
Gregorian telescope with the secondary (position, shape and size) still 
optimized for the 1.3 m Standard configuration: the main beam resolution is 
then equivalent to the 1.3 m Standard telescope. 

The average of the FWHM$_{\rm eff}$ in the relevant regions
(between $\sim -2.5$ and $\sim 2.5$ degrees 
for the 1.3m telescopes and between $\sim -2$ and $\sim 2$ degrees 
for the 1.5m telescopes)
are similar for Standard and Aplanatic configurations with same aperture
($\simeq 10$ arcmin for the $\simeq 1.5$ m telescopes and even better 
for the 1.75 m telescope).
On the other hand the FWHM$_{\rm eff}$ of beams located at angular
distances from the center 
roughly equal or larger than $\simeq 1^{\circ}$, 
where typical Planck (LFI and also HFI) feeds are located,
is somewhat better 
and also the spread of the FWHM$_{\rm eff}$'s of the different beams
is smaller for the Aplanatic configuration (Villa et~al.~1998b). 

\begin{figure}[h]
\centerline{\psfig{file=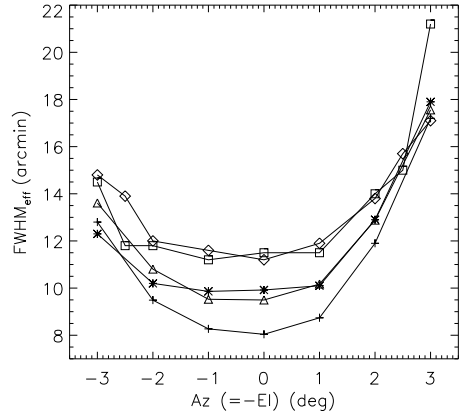,
width=7cm}}
\caption{FIGURE 4. Effective FWHM of beams located 
at different distance from the optical axis along a diagonal 
on the sky field of view, for the considered configurations:
the 1.3~m Standard telescope (diamonds), the 1.5~m Standard telescope
(triangles), the 1.75~m Standard telescope (crosses), the 1.3~m Aplanatic 
telescope (squares) and the 1.5~m Aplanatic telescope (asteriscs).}
\end{figure}

We find that the beam shapes are more regular for the Aplanatic configuration, 
and, although elliptical, closer to gaussian shapes,
due to the strong reduction of the coma. 
This could help also the reconstruction in flight of the beam pattern.

In addition, 
the Aplanatic configuration leads essentially unchanged the edge taper
at the bottom edge of the main reflector ($\sim -30$dB for the central
feed) compared to the Standard telescope, while it allows to 
improve the edge taper at the main reflector top edge ($\sim -40$dB for
the central feed), where the spillover radiation is not shielded. This will
most probably lead to an improvement of the top edge straylight. 
 
This preliminary study suggests that the Aplanatic configuration
can represent a significant improvement for the main beam properties 
compared to the Standard configuration, possibly decreasing 
the sidelobe contamination.
Further studies which include straylight, focal surface and feed positioning 
optimization, and mirror shapes, need to be done.

}

%

\bsk
\baselineskip = 9pt


{\references \ni REFERENCES
\ssk

\ref
Bersanelli, M., et al. 1996, ESA, COBRAS/SAMBA Report on the Phase A 
Study, D/SCI(96)3

\ref
Bersanelli, M., et al. 1998, Experimental Astronomy, in press

\ref
Bersanelli, M., Mandolesi, N. 1998, this Conference

\ref
Bond, R.J., Efstathiou, G., Tegmark, M. 1997, MNRAS 291, L33

\ref
Burigana, C., et al. 1997, Int. Rep. TeSRE/CNR 198/1997

\ref
Burigana, C., et al. 1998a, A\&ASS 130, 551

\ref
Burigana, C., et al. 1998b, this Conference

\ref 
Cavaliere, A., Danese, L., De~Zotti, G. 1979, A\&A 75, 322 

\ref
Delabrouille, J. 1998, A\&ASS 127, 555 

\ref
De Zotti, G., Toffolatti, L. 1998, this Conference


\ref
Dragone, C. 1978, B.S.T.J., Vol. 57, No. 7, 2663

\ref
Mandolesi, N., et al. 1997, Int. Rep. TeSRE/CNR 199/1997

\ref
Mandolesi, N., et al. 1998, Planck LFI,
A Proposal Submitted to the ESA. 


\ref
Mizuguchi, Y., Akagawa, M., Yokoi, H. 1978, Electronics \& Comm. in Japan, 
Vol. 61-B, No.~3,~58

\ref
Myers, S.T, Baker, J.E., Readhead, A.C.S., Leitch, E.M., Herbig, T. 1997, 
ApJ 485, 1

\ref
Pasian, F., Gispert, R. 1998, this Conference 
 
\ref 
Polegre, A.M., et al. 1998, this Conference

\ref
Puget, J.-L., et al. 1998, HFI for the Planck Mission,
A Proposal Submitted to the ESA.

\ref
Rephaeli, Y., 1995, ARA\&A 33, 541



\ref
Rush, W.V.T., et al. 1990, 
IEEE Trans. AP., Vol. 38, No. 8, 1141





\ref
Seiffert, M., et al. 1997, The Review of Scientific Instruments, submitted

\ref
Toffolatti, L., et al. 1998, MNRAS 297, 117

\ref
Villa, F., Bersanelli, M., Mandolesi, N. 1997, Int. Rep. TeSRE/CNR 188/1997

\ref
Villa, F., Bersanelli, M., Mandolesi, N. 1998a, Int. Rep. TeSRE/CNR 206/1998

\ref
Villa, F., Mandolesi, N., Burigana, C. 1998b, Int. Rep. TeSRE/CNR 221/1998

\ref
Wade, L.A., Levy, A.R. 1997, Cryocoolers, 9, 587


\ref 
Zaldarriaga, M., Spergel, D., Seljak, U. 1997, ApJ 488, 1

}                      

\end{document}